\documentclass[aps,prd,preprint,superscriptaddress,preprintnumbers,eqsecnum,nobibnotes,showpacs]{revtex4}
\usepackage{amsfonts,amssymb,amsmath,bm}
\usepackage{graphicx}
\usepackage{pstricks,color}

\newcommand{\be}{\begin{equation}}
\newcommand{\bea}{\begin{eqnarray}}
\newcommand{\ee}{\end{equation}}
\newcommand{\eea}{\end{eqnarray}}
\newcommand{\bpi}{\begin{picture}}
\newcommand{\bce}{\begin{center}}
\newcommand{\epi}{\end{picture}}
\newcommand{\ece}{\end{center}}

\def\chic#1{{\scriptscriptstyle #1}}

\def\gb{\bm{\Gamma}}

\begin{document}

\title{Indirect determination of the Kugo-Ojima function \\
from lattice data}

\author{A.~C. Aguilar}
\affiliation{Federal University of ABC, CCNH, 
Rua Santa Ad\'elia 166,  CEP 09210-170, Santo Andr\'e, Brazil.}
\author{D.~Binosi}
\affiliation{European Centre for Theoretical Studies in Nuclear
  Physics and Related Areas (ECT*), Villa Tambosi, Strada delle
  Tabarelle 286, I-38050 Villazzano (TN), Italy.}
\author{J. Papavassiliou}
\affiliation{Department of Theoretical Physics and IFIC, 
University of Valencia-CSIC,
E-46100, Valencia, Spain.}

\begin{abstract}

We study  the structure and  non-perturbative properties of  a special
Green's function, $u(q^2)$, whose infrared  behavior has traditionally served as
the  standard   criterion  for  the  realization   of  the  Kugo-Ojima
confinement mechanism.  It  turns out that, in the  Landau gauge, 
$u(q^2)$  can be determined  from a dynamical  equation, whose
main  ingredients are  the  gluon propagator  and  the ghost  dressing
function, integrated  over all physical  momenta.  Using as  input for
these two (infrared finite)  quantities recent lattice data, we obtain
an indirect determination of $u(q^2)$.  The results of this mixed
procedure are  in excellent agreement with those  found previously on
the  lattice, through  a  direct simulation  of  this function.   Most
importantly, in  the deep infrared the  function deviates considerably
from the  value associated with the realization  of the aforementioned
confinement scenario.   In addition, the dependence  of $u(q^2)$,
and  especially of  its value  at the  origin, on  the renormalization
point is  clearly established.  Some  of the possible  implications of
these results are briefly discussed.

\end{abstract}


\maketitle

\section{Introduction}

The problem of quark confinement and gluon screening is of central importance  
in QCD, and a large body of work has been dedicated to its understanding~\cite{Greensite:2003bk}.
Some of the most widely explored mechanisms attempting to explain how quarks confine
make concrete predictions about the non-perturbative behavior of 
the fundamental Green's functions of the theory.
For example, a central ingredient in the center vortex picture of confinement 
put forth by Cornwall  is the dynamical generation of a 
gluon mass~\cite{Cornwall:1982zr} through the well-known Schwinger mechanism~\cite{Schwinger:1951ex,Jackiw:1973tr,Farhi:1982vt},
implemented within the pinch technique (PT) framework~\cite{Cornwall:1982zr,Cornwall:1989gv,Binosi:2002ft}. 
In addition to taming the infrared divergences 
intrinsic to perturbation theory (Landau pole), this mass gives rise  
to a low energy effective theory~\cite{Cornwall:1974hz} which supports  
quantum solitons (center vortices), not present in the massless theory, 
whose condensation furnishes an  area law to  the fundamental representation  Wilson loop,
thus confining quarks~\cite{Brodsky:2008be}.
On the  other hand, the adjoint potential shows a
roughly linear  regime followed by string breaking  when the potential
energy is about $2m$, where $m$ is
the induced mass of the gluon \cite{Bernard:1981pg, Donoghue:1983fy}, 
corresponding to gluon screening~\cite{Bernard:1982my,Cornwall:1997ds}.
At the level of the two fundamental Green's functions of the theory, 
namely the gluon and ghost propagators, the predictions of the above picture 
are very definite: the gluon propagator is infrared finite 
(due to the generation of the gluon mass~\cite{Aguilar:2008xm}, 
whose phenomenological value
has been delimited in Ref.~\cite{Parisi:1980jy}), while, as has been shown recently~\cite{Aguilar:2008xm,Boucaud:2008ji},
in the Landau gauge the ghost 
remains massless, but with a finite dressing function 
(due to the saturation produced by the gluon mass)~\cite{ghostmass}.

An entirely different set of predictions is obtained within the 
Kugo-Ojima (KO) scenario, which also establishes a highly non-trivial link  
between confinement and the infrared behavior of some of the 
most fundamental Green's functions of QCD. 
In the KO confinement picture (in   covariant    gauges), the absence
of colored asymptotic states from  the physical spectrum of the theory
is   due   to   the   so-called   ``quartet mechanism''~\cite{Kugo:1979gm}.  
A  sufficient  condition  for  the realization of this mechanism
(and the meaningful definition of a conserved BRST charge) 
is that a certain correlation function, to be denoted by $u(q^2)$,
defined in Eq.~(\ref{KO-1}), should satisfy the condition $u(0)=-1$~\cite{KOdef}. 
In  addition, as first noted by  Kugo~\cite{Kugo:1995km}, in the
Landau gauge, $u(0)$ is  related to the  infrared behavior of the ghost dressing
function  $F(q^2)$ [see  Eq.~(\ref{ghostdress})] through  the identity
$F^{-1}(0)=1+u(0)$.  Therefore, the  KO confinement  scenario  predicts a
divergent ghost dressing function, and vice-versa.
Interestingly enough, the same prediction about $F^{-1}(0)$
is obtained when implementing the Gribov-Zwanziger (GZ) horizon condition~\cite{Gribov:1977wm,Zwanziger:1993dh}: 
in the IR region the ghost propagator diverges more rapidly 
than at tree-level~\cite{Zwan}. Furthermore, it has been also argued that the 
Landau gauge gluon propagator should vanish in the same limit~\cite{KOpred}. 
This alleged connection between confinement, the horizon condition, and an infrared ``enhanced'' 
ghost dressing function  has served as the theoretical cornerstone of the 
so called ``ghost-dominance'' picture of QCD~\cite{Fischer:2006ub}.

Turns   out    that   recent   large    volume   lattice   simulations~\cite{Cucchieri:2007md,Bogolubsky:2007ud} appear  
to be at  odds with
the original KO and GZ pictures  described above, 
at least as far as
their predictions about the infrared behavior of the Green's functions
are concerned~\cite{Dudal:2008sp}.  Specifically, various lattice studies, both in $SU(2)$
and  $SU(3)$, find  (in the  Landau  gauge) an  infrared finite  gluon
propagator~\cite{Alexandrou:2000ja}  and an infrared  finite (``non-enhanced'')  ghost dressing
function. Evidently, if taken  at face  value~\cite{grilatt}, these  results furnish
strong support for  the PT picture of dynamical  gluon mass generation
and the ensuing confinement mechanism.

It is perfectly clear that further detailed scrutiny from all possible angles must be 
implemented before reaching a definite conclusion on any of these issues. 
In this vein, it is natural to ask 
what one really knows about the KO function $u(q^2)$.  
Turns out that $u(q^2)$ 
has been studied directly on the 
lattice using the field-theoretic definition of $u(q^2)$ appearing in the KO formulation.
The first such study dates back to the work of 
Nakajima and Furui~\cite{Nakajima:1999dq}, who reported a 
value of $u(0)$ of about $-0.8$. More recently,   
Sternbeck~\cite{Sternbeck:2006rd} presented 
large-volume lattice simulations of the KO function (renormalized within the MOM-scheme). 
As can be plainly seen from Sternbeck's results (reproduced for convenience in Fig.~\ref{fig3-bis} of this article), $u(0)$ 
deviates appreciably from its KO value of $-1$; specifically, 
the function $u(q^2)$ saturates in the deep infrared around approximately  $-0.6$.
Interestingly enough, in a  recent article~\cite{Kondo:2009ug} Kondo gave a simple
 derivation of this same value, after appropriately modifying the KO construction 
in order to self-consistently accommodate the GZ horizon condition. 

Quite remarkably, in the (background) Landau gauge~\cite{Grassi:2004yq} the KO function coincides
with a certain auxiliary function, usually denoted by $G(q^2)$,   
which constitutes a crucial ingredient 
in the modern formulation of the PT by means of the   
Batalin-Vilkovisky (BV) quantization formalism~\cite{Batalin:1977pb}. 
Specifically, $G(q^2)$ is the form-factor multiplying $g_{\mu\nu}$ 
in the Lorentz decomposition of a 
special Green's function, denoted by $\Lambda_{\mu\nu}(q)$, which 
enters in all ``background-quantum'' identities~\cite{Grassi:1999tp,Binosi:2002ez}, {\it i.e.}, the infinite 
tower of non-trivial relations connecting the Green's functions of the 
background field method (BFM)~\cite{Abbott:1980hw} 
to the conventional ones (e.g. $R_{\xi}$ gauges). 
Notice also that $G(q^2)$ 
plays a prominent role in the new Schwinger-Dyson (SD) equations derived 
within the PT framework~\cite{Binosi:2007pi}, which, due to the special properties of 
the Green's functions involved, can be truncated in a manifestly gauge 
invariant way~\cite{Aguilar:2006gr}.

As has been shown in a recent article~\cite{Aguilar:2009nf}, 
one may derive a dynamical (SD-like) equation for $G(q^2)$, 
which, under mild assumptions,  
allows one to reconstruct $G(q^2)$ from the knowledge 
of the  gluon and ghost propagators. 
Specifically, $G(q^2)$ is determined by integrating over all virtual  
momenta ($k$) a kernel involving the product $\Delta(k) F(k+q)$. 
We emphasize that the aforementioned dynamical equation is not a simple relation 
of several Green's functions at some special isolated point; instead,  
the value obtained for $G(q^2)$ at any point (such as $q^2 =0$)
must be compatible with the behavior of $F$ and $\Delta$ 
in the entire range of their physical (euclidean) momenta. 
In particular, one must know their 
behavior not only in the IR, but also in the intermediate 
region of momenta (0.3-3 GeV), which appears rather difficult to obtain from 
SD studies~\cite{under}. 
This feature is very powerful, because it probes the details of the fundamental Green's functions 
over an extended range of momenta, rather then  just a single point.

In  the  present work,  we  
use the  available lattice data  on the gluon  and ghost propagator 
as input into the aforementioned dynamical  equation, 
thus obtaining an indirect determination of $G(q^2)$ 
in the entire range of available lattice momenta. 
Given the Landau gauge coincidence between $G(q^2)$ and $u(q^2)$, this procedure 
automatically determines the KO function as well.
This, in turn, permits us to 
obtain the  value of  the KO  parameter $u(0)$, as well as the GZ 
horizon  function, and  study their  dependence on  the renormalization
point $\mu$. 
Our analysis reveals an impressive self-consistency between the various 
ingredients entering into the calculation.
In particular, the results obtained through our combined method 
(SD using lattice data as input) are in excellent agreement 
with those of  \cite{Sternbeck:2006rd}, obtained through a direct simulation of 
the KO function.

The paper is organized as follows. In Section~\ref{review} we briefly introduce  
the BV framework for $SU(N)$ Yang-Mills theories, 
where the function $\Lambda_{\mu\nu}(q)$ appears naturally. 
Next, we review a number of relations where this function plays a key role: ({\it i}) 
the background-quantum identity relating the conventional and the BFM gluon propagators; 
({\it ii}) the relation between the ghost dressing function $F(q^2)$ and the $\Lambda_{\mu\nu}(q)$ 
form factors $G(q^2)$ and $L(q^2)$; ({\it iii}) we establish the crucial equality $u(q^2)= G(q^2)$; 
({\it iv}) the relation with the GZ horizon function. 
In addition, in the last subsection we discuss the renormalization of the 
 KO function and the resulting dependence on the renormalization point $\mu$, focusing particularly  
on how this latter dependence manifests itself within the MOM scheme.
The central results of this article are presented in Section~\ref{numana}. Specifically, 
the Lorentz decomposition of $\Lambda_{\mu\nu}(q)$ gives rise to two form-factors, 
the $G(q^2)$, which in the previous section has been identified 
with the KO function $u(q^2)$, and the $L(q^2)$, which has the 
particular property of vanishing in the deep IR.  
After establishing the dynamical equations governing $G(q^2)$ and  $L(q^2)$,  
we use the recent lattice data for the gluon and ghost propagators 
as input in these equations. 
In addition to the equations for $G(q^2)$ and $L(q^2)$, we consider 
the SD equation for the ghost, which is calibrated in order to be numerically 
compatible with the lattice data (at an impressive precision) simply by adjusting 
the gauge coupling to values that are slightly above the standard two-loop MOM prediction;
the obtained value of the coupling is then used into the equations for $G(q^2)$ and $L(q^2)$. 
We use the multiplicative renormalizability of the gluon and ghost propagators in order to rescale the 
lattice data to different values of the renormalization point. Even though this procedure has an 
intrinsic limitation set by the relatively short reach of the available data into the UV,   
it amply demonstrates that $u(q^2)$ depends non-trivially on $\mu$, in excellent agreement with the 
observation established in \cite{Sternbeck:2006rd}. 
Finally, in Section~\ref{concl} we present our conclusions.

\section{\label{review} Connecting the Kugo-Ojima and G functions}

As already mentioned in the Introduction, in the Landau gauge 
the KO function  $u(q^2)$ may be shown to be identical to the function $G(q^2)$, which appears 
in several formal contexts.     
In this section we first formulate Yang-Mills theories in the BV framework, 
which allows the derivation of a tower of identities, whose common ingredient is the 
function $G(q^2)$. Then we will show why $u(q^2)=G(q^2)$, and will review the 
connection between $u(q^2)$ and the GZ horizon. 
This main purpose of this section is to serve as a reminder and to  
bring together various seemingly 
disjoint pieces of information. For specific 
details on each topic the reader is referred to the corresponding extensive literature.

\subsection{Batalin-Vilkoviski formalism}

In the BV formulation of Yang-Mills theories~\cite{Batalin:1977pb}, 
one starts by introducing certain sources (called anti-fields in what follows) 
that describe the renormalization of composite operators; the latter class 
of operator is in fact bound to appear in such theories due to the non-linearity 
of the BRST transformation of the elementary fields. In much the same way, 
the quantization of the theory in a background field type of gauge requires, in addition 
to the aforementioned anti-fields, the introduction of new sources which couple to 
the BRST variation of the background fields~\cite{Grassi:1999tp}. 
These sources are sufficient for implementing the full set of symmetries of a 
non-Abelian theory at the quantum level, and in the case of quarkless $SU(N)$ QCD, lead to the master equation
\be
\int\!d^4x\left[\frac{\delta\Gamma}{\delta A^{*m}_\mu}\frac{\delta\Gamma}{\delta A^{m}_\mu}+\frac{\delta\Gamma}{\delta c^{*m}}\frac{\delta\Gamma}{\delta c^m}+B^m\frac{\delta\Gamma}{\delta\bar c^m}+\Omega^m_\mu\left(\frac{\delta\Gamma}{\delta \widehat{A}^m_\mu}-\frac{\delta\Gamma}{\delta A^m_\mu}
\right)\right]=0.
\label{me}
\ee
In the formula above, $\Gamma$ is the effective action, $A^*$ and $c^*$ the gluon and ghost anti-fields, $\widehat{A}$ 
is the gluon background field, and $\Omega$ the corresponding background source; finally $B$ denotes 
the Nakanishi-Lautrup multiplier for the gauge fixing condition.

To determine the complete algebraic structure of the theory we need two additional equations. 
The first one is the Faddeev-Popov equation, that controls the result of the contraction 
of an anti-field  leg with the corresponding momenta. In position space, it reads
\be
\frac{\delta\Gamma}{\delta \bar c^m}+\left(\widehat{\cal D}^\mu\frac{\delta\Gamma}{\delta A^*_\mu}\right)^m-\left({\cal D}^\mu\Omega_\mu\right)^m-gf^{mrs}\widehat{A}^r_\mu\Omega^\mu_s=0,
\label{FPE}
\ee
where
$({\cal D}^\mu \Phi)^m=\partial^\mu \Phi^m+gf^{mnr}A^n_\mu \Phi^r$ [in the case of $(\widehat{\cal D}^\mu \Phi)^m$ replace the gluon field $A$ with a background gluon field $\widehat{A}$].
The second one is the anti-ghost equation formulated in  the background field Landau gauge, which 
reads~\cite{Grassi:2004yq}
\be
\frac{\delta\Gamma}{\delta c^m}-\left(\widehat{\cal D}^\mu \frac{\delta \Gamma}{\delta \Omega_\mu}\right)^m-\left({\cal D}^\mu A^*_\mu\right)^m-gf^{mnr}c^{*n}c^r
+gf^{mnr}\frac{\delta\Gamma}{\delta B^n}\bar{c}^r=0,
\label{AGE}
\ee
This equation fully constrains the dynamics of the ghost field $c$, and implies that the latter 
will not get an independent renormalization constant.  The local form of the anti-ghost equation (\ref{AGE}) 
is only valid when choosing the background Landau gauge condition $(\widehat{\cal D}^\mu A_\mu)^m=0$; 
in the usual Landau gauge, $\partial^\mu A_\mu^m=0$, an integrated version of this equation is available. 
In fact, even though the results that follow will be derived for convenience in the background Landau gauge, they are valid also in the conventional Landau gauge of the $R_\xi$.

\begin{figure}[!t]
\includegraphics[width=10cm]{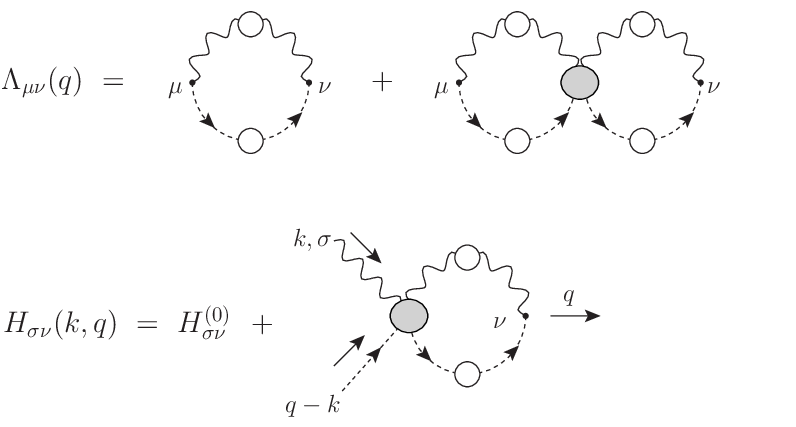}
\caption{Diagrammatic representation of the functions $\Lambda$ and $H$.}
\label{fig:Lambda-H}
\end{figure}

Now, differentiation of the functional (\ref{me}) with respect to a combination of fields containing 
at least one ghost field or two ghost fields and one anti-field (and setting the fields and sources 
to zero afterwards) will provide the Slavnov-Taylor identities of the theory. Differentiation 
with respect to a background source and background or quantum fields will provide, instead, the 
so called background-quantum identities~\cite{Grassi:1999tp,Binosi:2002ez}, which relate 1PI Green's functions 
involving background fields with those involving quantum fields. 
Finally, differentiation of (\ref{FPE}) and (\ref{AGE}) with respect to fields and anti-fields or 
background sources give rise to relation among the different auxiliary ghost functions appearing in the theory.

The important point is that, when carrying out these differentiations, the following function appears (Fig.~\ref{fig:Lambda-H})
\bea
i\Lambda_{\mu \nu}(q) &=&\Gamma_{\Omega_\mu A^*_\nu}(q)\nonumber \\
&=&g^2C_A
\int_k H^{(0)}_{\mu\rho}
D(k+q)\Delta^{\rho\sigma}(k)\, H_{\sigma\nu}(k,q),
\nonumber \\
&=&i g_{\mu\nu} G(q^2) + i\frac{q_{\mu}q_{\nu}}{q^2} L(q^2),
\label{LDec}
\eea
and (in $d$-dimensions)
\be
G(q^2) = \frac{1}{(d-1)q^2} \left(q^2 \Lambda_{\mu}^{\mu} - q^{\mu}q^{\nu}\Lambda_{\mu\nu} 
\right),\qquad L(q^2) = \frac{1}{(d-1)q^2} \left(d q^{\mu}q^{\nu}\Lambda_{\mu\nu} - q^2\Lambda_{\mu}^{\mu} \right).
\label{s3}
\ee 
In the equations above, the color factor $\delta^{mn}$ has been factored out (as always in what follows), 
$C_{\rm {A}}$ represents the Casimir eigenvalue of the adjoint representation
[$C_{\rm {A}}=N$ for $SU(N)$], 
and \mbox{$\int_{k}\equiv\mu^{2\varepsilon}(2\pi)^{-d}\int\!d^d k$}, 
with $d=4-\epsilon$ the dimension of space-time. 
$\Delta_{\mu\nu}$ and $D$ represents the gluon and ghost propagator respectively, defined as
\bea
\Delta_{\mu\nu}(q)&=&-i\left[ P_{\mu\nu}(q)\Delta(q^2) +\xi\frac{q_\mu q_\nu}{q^4}\right],
\label{prop_cov}\\
D(q^2)&=& \frac{iF(q^2)}{q^2},
\label{ghostdress}
\eea 
where $\xi$ denotes the gauge-fixing parameter, and 
\mbox{$P_{\mu\nu}(q)= g_{\mu\nu} - q_\mu q_\nu /q^2$}
is the usual transverse projector; notice that $\Delta^{-1}(q^2) = q^2 + i \Pi(q^2)$, 
with  $\Pi_{\mu\nu}(q)=P_{\mu\nu}(q)\Pi(q^2)$ the gluon self-energy. $F(q^2)$ is the so called ghost dressing function.
Finally, the function $H_{\mu\nu}(k,q)$ (see  Fig.~\ref{fig:Lambda-H} again)
is in fact a familiar object, since 
it appears in the all-order Slavnov-Taylor identity
satisfied by the standard  three-gluon vertex~\cite{Marciano:1977su}
\bea
q^\alpha\Gamma_{\alpha\mu\nu}(q,k_1,k_2)&=& iF(q^2)\Delta^{-1}(k_1^2)P_\mu^\rho(k_1)H_{\nu\rho}(k_2,k_1)\nonumber \\
&-& iF(q^2)\Delta^{-1}(k_2^2)P_\nu^\rho(k_2)H_{\mu\rho}(k_1,k_2).
\eea
It is also related to the full gluon-ghost vertex $\gb_{\mu}(k,q)$ by the identity
\be
q^\nu H_{\mu\nu}(k,q)=-i\gb_{\mu}(k,q).
\label{qH}
\ee
At tree-level, $H_{\mu\nu}^{(0)} = ig_{\mu\nu}$ and $\gb^{(0)}_{\mu}(k,q)=\Gamma_\mu(k,q)=-q_\mu$.

\subsection {Background-quantum identities}

The first identities where the function $\Lambda_{\mu\nu}$ appears 
are the so-called background-quantum identities, {\it i.e.}, the infinite tower of non-trivial relations connecting the 
BFM Green's functions to the conventional ones~\cite{Grassi:1999tp,Binosi:2002ez}.
Consider, for example, the result of differentiating the functional (\ref{me}) with respect to 
a background source and a background gluon, on the one hand, and a background source and a quantum gluon, on the other, one obtains 
two equations, 
\bea
\Gamma_{\widehat{A}_\mu    A_\nu}(q)&=&\left[g_{\mu\rho}
+  \Lambda_{\mu\rho}(q)\right]\Gamma_{A^\rho A_\nu}(q),\nonumber \\
\Gamma_{\widehat{A}_\mu\widehat{A}_\nu}(q)&=&\left[g_{\mu\rho}
+  \Lambda_{\mu\rho}(q)\right]\Gamma_{A^\rho
\widehat A_\nu}(q).
\label{twobqi}
\eea
Using the transversality of the gluon two-point function, these two equations can then be 
appropriately combined to yield the important identity 
\be
\Gamma_{\widehat{A}_\mu  \widehat{A}_\nu}(q)=\left[1+G(q^2)\right]^2\Gamma_{A_\mu A_\nu}(q),
\ee
or, in terms of propagators
\be
\widehat\Delta^{-1}(q^2)=\left[1+G(q^2)\right]^2\Delta^{-1}(q^2).
\label{DeltaBQI}
\ee

The quantity $\widehat\Delta(q^2)$ appearing on the left-hand side of the above equation
captures the running of the QCD $\beta$ function, exactly as happens with the QED vacuum polarization; this is a fundamental property of the 
BFM gluon self-energy, valid for every value of the (quantum) gauge-fixing parameter~\cite{Abbott:1980hw}.
This can be easily checked to lowest order, where Eqs.~(\ref{LDec})  and~(\ref{s3}) give (in the Landau gauge)
\bea
1+G(q^2) &=& 1 +\frac{9}{4}
\frac{C_{\rm {A}}g^2}{48\pi^2}\ln\left(\frac{q^2}{\mu^2}\right),\nonumber \\
\Delta^{-1}(q^2) &=& q^2 \left[1+\frac{13}{2}
\frac{C_{\rm {A}}g^2}{48\pi^2}\ln\left(\frac{q^2}{\mu^2}\right)\right],
\label{pert_gluon}
\eea
and thus
\be
\widehat\Delta^{-1}(q^2) = q^2 \left[1+ b g^2 \ln\left(\frac{q^2}{\mu^2}\right)\right].
\ee
where $b=11 C_{\rm {A}}/48\pi^2$ is the first coefficient in the QCD $\beta$ function.  
Eq.~(\ref{DeltaBQI})
plays a central role in the derivation of a new set of SDEs~\cite{Binosi:2007pi} that can be truncated in a manifestly gauge invariant way~\cite{Aguilar:2006gr}. 

Let us conclude this subsection by noticing that in more general identities the function $L(q^2)$ is also relevant. Consider {\it e.g.}, 
the identity relating the three-gluon proper vertices. One then has
\bea
\Gamma_{\widehat{A}_\mu A_\alpha A_\beta}(k_1,k_2)&=& \left[g^\nu_\mu + \Lambda^\nu_\mu(q)\right]
\Gamma_{A_\nu A_\alpha A_\beta}(k_1,k_2)+\cdots\nonumber \\
&=&\left[1+G(q^2)\right]\Gamma_{A_\mu A_\alpha A_\beta}(k_1,k_2)+\frac{q_\mu q^\nu}{q^2}L(q^2)\Gamma_{A_\nu A_\alpha A_\beta}(k_1,k_2)+\cdots,\qquad
\eea
where the omitted terms involve  
other auxiliary Green's functions (see~\cite{Grassi:1999tp,Binosi:2002ez}), irrelevant to our discussion.

\subsection{Two-point ghost sector}

Let us now consider the two-point functions. Differentiating the ghost equation (\ref{FPE}) with respect to a ghost field and a background source we get the relations
\bea
\Gamma_{c\bar c}(q)&=&-iq^\nu\Gamma_{c A^{*}_\nu}(q),\nonumber \\
\Gamma_{\bar c\Omega_\mu}(q)&=&q_\mu+q^\nu\Lambda_{\mu\nu}(q).
\label{FPE_ghost}
\eea
On the other hand, differentiating the anti-ghost equation~(\ref{AGE}) with respect to a gluon anti-field and an anti-ghost, one gets
\bea
\Gamma_{c A^{*}_\nu}(q)&=&q_\nu+q^\mu\Lambda_{\mu\nu}(q),
\nonumber \\
\Gamma_{c\bar c}(q)&=&-iq^\mu\Gamma_{\bar c\Omega_\mu}(q).
\label{AGE-1}
\eea

Next,  contracting the first equation in~(\ref{AGE-1}) with $q^\nu$, and making use of  the first equation in~(\ref{FPE_ghost}), 
we see that the dynamics of the ghost sector is entirely captured by the $\Lambda_{\mu\nu}$ auxiliary function, since
\be
i\Gamma_{c\bar c}(q)=q^2+q^\mu q^\nu\Lambda_{\mu\nu}(q).
\label{funrelbv}
\ee
Introducing  the Lorentz decompositions
\be
\Gamma_{cA^*_\mu}(q)=q_\mu C(q^2),\qquad \Gamma_{\bar c \Omega_\mu}(q)=q_\mu E(q^2),
\ee
we find that Eq.~(\ref{funrelbv}) together with the last equation of~(\ref{FPE_ghost}) and~(\ref{AGE-1}) give 
the identities~\cite{Kugo:1995km,Grassi:2004yq}
\bea
C(q^2)& =& E(q^2)\ =\ F^{-1}(q^2),\nonumber \\
F^{-1}(q^2)&=&1+G(q^2)+L(q^2).
\label{ids}
\eea

Finally, recalling that the dimension of the gluon anti-field $A^*$ is three, 
while the dimension of the $\Omega$ source is one, power counting  shows that  
({\it i}) all functions appearing in Eqs.~(\ref{FPE_ghost}) and~(\ref{AGE-1}) are divergent, 
and ({\it ii}) the divergent part of $\Lambda_{\mu\nu}(q)$ can be 
proportional to $g_{\mu\nu}$ only~\cite{Grassi:2004yq,Aguilar:2009nf}.

\subsection{The (background) Landau gauge equality between $u(q^2)$ and $G(q^2)$}

\begin{figure}[!t]
\begin{center}
\includegraphics{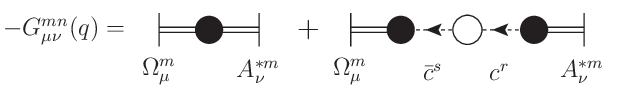}
\end{center}
\caption{Connected components contributing to the function $G^{mn}_{\mu\nu}(q)$.}
\label{connect}
\end{figure}

A crucial ingredient to our analysis is the equality between the KO function and the $G(q^2)$, in the (background) Landau gauge~\cite{Grassi:2004yq}.
To see this, we start with the following (Euclidean) two-point function of composite operators
\be
\int\!d^4x\ \mathrm{e}^{-iq\cdot(x-y)}\langle T\big[\left({\cal D}_\mu c\right)_x^m\left(f^{nrs}A^n_\nu\bar c^s\right)_y\big]\rangle=P_{\mu\nu}(q)\delta^{mn}u(q^2),
\label{KO-1}
\ee
which, due to the identity
\be
\int\!d^4x\ \mathrm{e}^{-iq\cdot(x-y)}\langle T\big[\left({\cal D}_\mu c\right)_x^m\bar c^n\big]\rangle=-i\frac{q_\mu}{q^2}\delta^{mn}
\ee
can be related to the function $\langle\big[\left({\cal D}_\mu c\right)_x^m\left({\cal D}_\mu \bar c\right)_y^n\big]\rangle$, through
\be
\int\!d^4x\ \mathrm{e}^{-iq\cdot(x-y)}\langle T\big[\left({\cal D}_\mu c\right)_x^m\left({\cal D}_\mu \bar c\right)_y^n\big]\rangle=
-\frac{q_\mu q_\nu}{q^2}\delta^{mn}+P_{\mu\nu}(q)\delta^{mn}u(q^2).
\label{id-1}
\ee
On the other hand, observe that in the background Landau gauge the function appearing on the lhs of the above equation is precisely given by
\be
-{\cal G}^{mn}_{\mu\nu}(q)=\frac{\delta^2 W}{\delta \Omega^{m}_\mu\delta A^{*n}_\nu},
\ee
where $W$ is the generator of the connected Green's functions, and the two connected diagrams contributing to  ${\cal G}_{\mu\nu}$ are shown in Fig.~\ref{connect}. Factoring out the color structure and making use of the identities~(\ref{ids}) one has
\bea
-i{\cal G}_{\mu\nu}(q)&=&
\Lambda_{\mu\nu}(q)+ \Gamma_{\Omega_\mu\bar c}(q)D(q^2)\Gamma_{A^*_\nu c}(q)\nonumber \\
&=&-\frac{q_\mu q_\nu}{q^2}+P_{\mu\nu}(q)G(q^2).
\label{id-2}
\eea
Passing to the Euclidean formulation, and comparing with Eq.~(\ref{id-1}), we then arrive at the important equality  
\be
u(q^2)=G(q^2).
\label{KO-G}
\ee
Then, the usual KO confinement criterion may be equivalently cast in the form: $1+G(0)=0$. 
Evidently, if $L(0)=0$ [see discussion after Eq.~(\ref{LGr})], then 
from the identity (\ref{ids}) follows that if the KO criterion is satisfied then the ghost dressing function 
diverges in the IR.

\subsection{Gribov-Zwanziger horizon}

In order to avoid Gribov copies~\cite{Gribov:1977wm}, in the GZ formulation of Yang-Mills theories
the partition function assumes the form  (in  $d$-dimensional Euclidean space)~\cite{Zwanziger:1993dh}
\be
Z_\gamma=\int[dA]\delta(\partial^\mu A_\mu)\mathrm{Det} M\exp\left\{-S_\mathrm{YM}+\gamma\int\!d^dx\,h(x)\right\},
\ee
where $S_\mathrm{YM}$ is the Yang-Mills action, $M=-\partial_\mu{\cal D}^\mu$ is the Faddeev-Popov operator, 
and the functional $h(x)=h[A](x)$ is the so-called GZ horizon function given by
\be
h(x)=-\int\!d^dy\,gf^{amr}A^m_\mu(x)(M^{-1})^{rs}(x,y)gf^{asn}A^n_\nu(y);
\ee
thus, the action corresponding to the partition function above clearly contains a non-local term.
The GZ parameter, $\gamma$, is determined through the so-called {\it horizon condition}, which for  $SU(N)$ assumes the form
\be
\left\langle h(x)\right\rangle_\gamma=d(N^2-1).
\label{sccon}
\ee
This condition can be rewritten in terms of the vev of the GZ horizon function if we integrate both sides over $d^dx$.  In this way we get
\bea
\left\langle h(0)\right\rangle_\gamma&\equiv&\frac1{V_d}\frac{\partial}{\partial\gamma}\ln Z_\gamma\nonumber \\
&=&\frac1{V_d}\int\!d^dx\left\langle h(x)\right\rangle_\gamma=d(N^2-1).
\eea
On the other hand, assuming that $\gamma$ is small, one can expand in powers of $\gamma$; retaining the first order only, one gets 
\be
\left\langle h(0)\right\rangle_\gamma\simeq\left\langle h(0)\right\rangle_{\gamma=0}+{\cal O}(\gamma).
\label{approx}
\ee

The right-hand side of the above equation can be related to the trace of 
the following Green's function (Euclidean space)  
\be
{\cal H}^{mn}_{\mu\nu}(q)=\delta^{mn}\left[P^\rho_\mu(q)+\frac{q_\mu q^\rho}{q^2} F(q^2)\right]
\Lambda_{\rho\nu}(q), 
\ee
in the limit $q^2\to0$ ~\cite{Kondo:2009ug}. 
Specifically, 
\bea
\left\langle h(0)\right\rangle_{\gamma=0}&=&\frac1{V_d}\int\!d^dx\left\langle h(x)\right\rangle_{\gamma=0}\ = \
 -\lim_{q^2\to0}\mathrm{Tr}\left\{{\cal H}^{mn}_{\mu\nu}(q)\right\}\nonumber \\
&=&-(N^2-1)\left\{(d-1)G(0)+F(0)\left[G(0)+L(0)\right]\right\}.
\label{res-1}
\eea

This result allows to rewrite the GZ horizon condition in terms of $G(0)$ (and therefore of the KO parameter $u(0)$); 
this will, in turn,  restrict the allowed values of $u(0)$. 
In the limit of vanishing Gribov parameter, one can use the result of (\ref{res-1}) 
to solve the horizon condition~(\ref{sccon}), in the approximation~(\ref{approx}); 
if $L(0)=0$, one finds (in $d=4$) the following value of the KO parameter~\cite{Kondo:2009ug}
\be
u(0)=G(0)=-\frac23, 
\label{new-pred}
\ee
which is very close to that obtained directly from the lattice ~\cite{Sternbeck:2006rd}, and, as we will see in the next section, from 
our independent analysis.

\subsection{Renormalization of  $u(q^2)$: the MOM scheme and the associated $\mu$-dependence}

Before entering into the specifics of the KO function, let us briefly  
recall some basic facts about renormalization.  
In general, Green's functions in $d=4$
must undergo renormalization.
The renormalization procedure renders the renormalized quantities  
UV finite, introducing at the same time a dependence on the 
renormalization point, denoted in general by  $\mu$. This dependence, usually referred to as ``$\mu$-dependence'',  
imposes non-trivial 
constraints on the asymptotic behavior of Green's functions, controlled by the 
renormalization group , and most concretely by the renormalization group equation corresponding 
to a given Green's function~\cite{Cheng:1985bj}.   
Specifically, a Green's function with $n$ incoming fields $\phi$, to be denoted (in momentum space) 
by $\Gamma^{(n)} (p_i,g,\mu)$, where $g$ is the coupling constant, satisfies for asymptotically large momenta
\be
\left(\mu \frac{\partial}{\partial \mu}  + \beta \frac{\partial}{\partial g} - n \gamma \right) \Gamma^{(n)}(p_i,g,\mu) = 0 \,,
\label{RGE}
\ee
where $\beta = \mu ({\partial g}/{\partial \mu})$,  
and $\gamma$ is the so called ``anomalous dimension'' of the 
field $\phi$, defined as $\gamma = \mu ({\partial Z_{\phi}}/{\partial \mu})$.
If the  Green's function under consideration contains composite operators ({\it i.e.}, $\phi^2(x)$), then 
Eq.~(\ref{RGE}) must be appropriately modified (see, for example,~\cite{Collins:1984xc}).

Note that the $\mu$-dependence infests also Green's functions that are UV finite, {\it i.e.}, they do not need 
explicit subtraction to be rendered finite ({\it i.e.}, no new counter-terms need be introduced) 
For example, in the $(\phi^4)_4$ theory,  all Green's functions with $n>4$ are UV finite, but depend 
in general on $\mu$.
In this case, the $\mu$-dependence enters 
through the dependence of the Green's functions on the propagators and vertices; 
since the latter depend explicitly on $\mu$,  $\Gamma^{(n)}$ (with $n>4$) develops 
a $\mu$-dependence through the higher loop corrections.

To be sure, a given Green's function may be renormalized at a fixed value of the incoming momenta, 
such as  $p_i^2=m_i^2$ (``on shell'' scheme), in which case there is no {\it manifest} dependence 
on $\mu$. 
Instead, within renormalization schemes such as the ${\overline{\rm MS}}$ or the ``MOM'', 
the intrinsic $\mu$-dependence of Green's functions becomes manifest. Note also an additional important point: 
the exact functional dependence of a given Green's function on $\mu$ changes from one renormalization scheme to another; 
for example, the $\mu$-dependence within the ${\overline{\rm MS}}$ does not coincide with that of the MOM. 

Turning now to $u(q^2)$, it is obvious from naive power-counting that 
it diverges logarithmically as the UV cutoff is taken to infinity. 
It is instructive to compute  $u(q^2)$ (in the Landau gauge) at one loop, to be denoted by $u^{[1]}(q^2)$. 
To that end we compute the integral in Eq.~(\ref{LDec}) at one loop, project out its $G(q^2)$ component using Eq.~(\ref{s3}),
and finally employ the crucial equality $u(q^2)=G(q^2)$ of Eq.~(\ref{KO-G}). 
To avoid IR divergences,  we introduce a hard gluon mass $m$ (in the next section 
this will be done properly, using 
the IR-finite gluon propagator obtained from the lattice). Then, a straightforward calculation 
yields (setting $z\equiv q^2/m^2$)
\be
u^{[1]}(z,\Lambda^2) = - \frac{3\alpha_s}{16 \pi} f(z,\Lambda^2)\,,
\label{u1x}
\ee
with
\be
f(z,\Lambda^2) = 3\ln \left(\frac{\Lambda^2}{m^2}\right) +  
\frac{19}{6} -\frac{1}{3z} - \left[3 + \frac{3}{z} - \frac{1}{3z^2} \right]\ln (1+z) 
+ \frac{z}{3}\ln \left(\frac{1+z}{z}\right) \,,
\label{fx}
\ee
where $\Lambda$ is the UV cutoff. Taking the limit of $z \to 0$ (expanding the logs), one finds that 
\be
u^{[1]}(0,\Lambda^2) = - \frac{9\alpha_s}{16 \pi}\ln \left(\frac{\Lambda^2}{m^2}\right) \,.
\ee
It is clear that a subtraction is sufficient to render $u^{[1]}(z,\Lambda^2)$ finite. 

In the present work 
we will use the MOM scheme to renormalize the pertinent Green's functions, and will impose the necessary normalization conditions 
in the deep UV (or, at least, as far into the UV as permitted by the lattice data), where perturbation theory is reliable, 
and let the non-perturbative dynamics (captured by the lattice, the SDE, etc) determine 
what the IR behavior of the Green's functions will be.
In fact, this latter renormalization procedure has been employed in  
practically all recent lattice studies; therefore, in order to be able to use self-consistently lattice data
as input to our SDE equations, and compare meaningfully our results with those of ~\cite{Sternbeck:2006rd}, 
we have to use the MOM scheme,  subject to an important constraint related with the 
preservation of the second identity in (\ref{ids}), as discussed in detail in subsection (\ref{numana} C).

Specifically, at the level of the one-loop example that we are considering in this subsection, the aforementioned 
constraint amounts 
to the statement that, due the validity of (\ref{ids})  and the fact that the function $L(z)$ does {\it not} 
vanish identically, 
one {\it cannot} impose tha standard MOM condition {\it simultaneously} on both the one-loop dressing 
function, $F^{[1]}_{\chic R}(z,\mu^2)$, and the KO function  $u^{[1]}_{\chic R}(z,\mu^2)$. In other words, one cannot 
have at the same time \mbox{$F^{[1]}_{\chic R}(z=\mu^2,\mu^2)= 1$} and $u^{[1]}_{\chic R}(z=\mu^2,\mu^2)=0$.

To see this explicitly, let us compute $L(z)$ at one-loop, to be denoted by $L^{[1]}(z)$, under the same assumptions employed before for $u^{[1]}(z)$;
it is straighforward to obtain the finite result 
\be
L^{[1]}(z) = \frac{\alpha_s}{4\pi}  \left\{
\frac{1}{z^2}\left[\frac{z^2}{2} - m^2 z +  m^4 \ln\left(1+\frac{z}{m^2}\right)\right]  
+ \frac{z}{m^2} \ln\left(1+\frac{m^2}{z}\right) \right\}.
\label{Lmx}
\ee
Note that $L^{[1]}(0)=0$; however, for any other finite value of $z$, $L^{[1]}(z) \neq 0 $.  
Given that (\ref{ids}) must remain valid, i.e. 
\be
\{F^{[1]}_{\chic R}(z,\mu^2)\}^{-1} = 1 +  u^{[1]}_{\chic R}(z,\mu^2) + L^{[1]}(z) \,,
\label{fsr}
\ee
it is clear that if one renormalizes the ghost according to the standard MOM prescription, $F^{[1]}(z=\mu^2,\mu^2)= 1$, then from  (\ref{fsr}) follows that 
\be
u^{[1]}_{\chic R}(z\!=\!\mu^2,\mu^2) = - L^{[1]}(\mu^2),
\label{fsrm}
\ee
which is the appropriate normalization condition for $u^{[1]}_{\chic R}$; of course,  $L^{[1]}(\mu^2)\neq 0$.  

\begin{figure}[!t]
\includegraphics[scale=0.8]{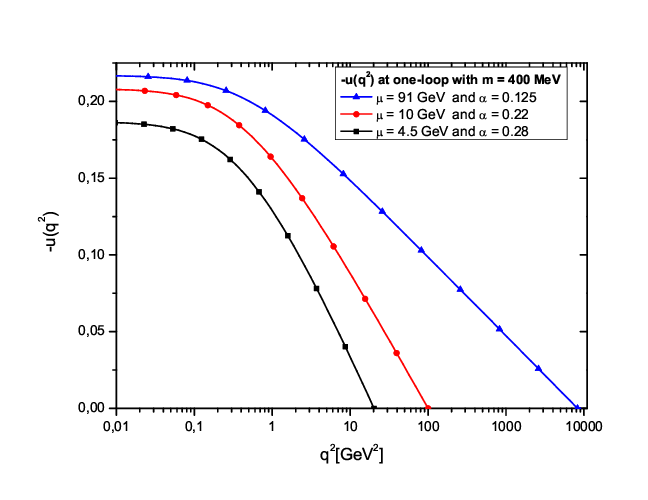}
\caption{The $\mu$-dependence of $u(q^2)$ at one loop, under the normalization condition of (\ref{fsrm}).}
\label{pmdol}
\end{figure}

The exact form of $u^{[1]}_{\chic R}(z,\mu^2)$ satisfying the normalization condition (\ref{fsrm}) is given by 
\be
u^{[1]}_{\chic R}(z,\mu^2) = - \frac{3\alpha_s}{16 \pi} f_{\chic R}(z,\mu^2) - L^{[1]}(\mu^2)\,,
\ee
where $f_{\chic R}(q^2,\mu^2) = f(q^2,\Lambda^2) - f(\mu^2,\Lambda^2)$, namely (setting $t \equiv \mu^2/m^2$)
\bea
f_{\chic R}(z,t) &=& 
-\frac{1}{3z} - \left[3 + \frac{3}{z} - \frac{1}{3z^2} \right]\ln (1+z) 
+ \frac{z}{3}\ln \left(\frac{1+z}{z}\right)
\nonumber\\
&& +\frac{1}{3t} + \left[3 + \frac{3}{t} + \frac{1}{3t^2} \right]\ln (1+t) 
- \frac{t}{3}\ln \left(\frac{1+t}{t}\right)\,.
\label{frx}
\eea
Evidently, $f_{\chic R}(t,t)=0$, or, equivalently $u(q^2\!\!=\!\!\mu^2,\mu^2) = - L^{[1]}(\mu^2)$, a required by (\ref{fsrm}).
The $\mu$-dependence induced to $u(q^2)$ after imposing the renormalization condition given in  (\ref{fsrm}) 
is shown in Fig.\ref{pmdol}. 

\section{\label{numana} Extracting the Kugo-Ojima function from the lattice}

In this section we study the behavior of the function $G(q^2)$ by using the available lattice data on the gluon and ghost propagators. 
Specifically, we will first write down the dynamical equations that govern the functions $G(q^2)$ and $L(q^2)$, 
which involve both $\Delta(q^2)$ and $D(q^2)$, and using as an input the lattice results for these propagators  
we will obtain an indirect determination of $G(q^2)$. 
Of course, by virtue of the fundamental equality $u(q^2)=G(q^2)$, any information on $G(q^2)$ translates automatically to the Kugo-Ojima function.

\subsection{Dynamical equations}

\begin{figure}[!t]
\includegraphics[width=11cm]{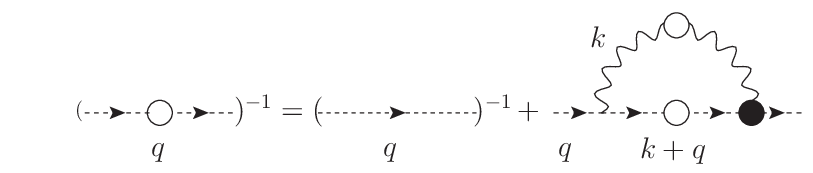}
\caption{The SDE satisfied by the ghost propagator.}
\label{ghostSDE}
\end{figure}

The dynamical equations governing $G(q^2)$ and $L(q^2)$ 
may be obtained directly from the defining equation (\ref{LDec}),  
by appropriately contracting it and taking its trace.
Specifically, from Eq.~(\ref{LDec}) one has~\cite{Aguilar:2009nf} (in $d$-dimensions, and setting $G(q^2) = u(q^2)$)
\bea
u(q^2) &=& \frac{g^2 C_{\rm {A}}}{d-1}
\left[ 
\int_k \Delta^{\rho\sigma}(k)\, H_{\sigma\rho}(k,q) D(k+q)
+i
\frac{1}{q^2} \int_k\! q^{\rho} \Delta_{\rho\sigma}(k)\, {\gb}^{\sigma}(k,q) D(k+q)
\right]\!,
\nonumber\\
L(q^2) &=& -\frac{g^2 C_{\rm {A}}}{d-1}
\left[i
\frac{d}{q^2} \int_k\! q^{\rho} \Delta_{\rho\sigma}(k)\, {\gb}^{\sigma}(k,q) D(k+q)
\!+\!\! \int_k\! \Delta^{\rho\sigma}(k)\, H_{\sigma\rho}(k,q) D(k+q)
\right]\!.
\label{s4}
\eea
Adding the above equations by parts, and employing (\ref{qH}), 
one may easily demonstrate~\cite{Aguilar:2009nf}, in the Landau gauge, the validity of
the second identity in Eq.~(\ref{ids}), given that the standard SD equation for the  ghost propagator (Fig.~\ref{ghostSDE}) 
reads 
\be
iD^{-1}(q^2) = q^2 +i g^2 C_{\rm {A}}  \int_k
\Gamma^{\mu}\Delta_{\mu\nu}(k)\gb^{\nu}(k,q) D(q+k).
\label{SDgh}
\ee

The two equations in (\ref{s4}) 
involve five basic ingredients:  the two-point functions $\Delta(q^2)$ 
and $D(q^2)$, the vertex functions $\gb_\mu(k,q)$ and $H_{\mu\nu}(q,k)$, and, eventually, 
the value of the 
(renormalized) coupling $g^2$, at different renormalization points.
Knowledge (direct or indirect)  of these ingredients (for example from the lattice) would, in turn,
determine fully the functions $u(q^2)$ and $L(q^2)$. 

\subsection{The vertices}

Let us begin with  $\gb_\mu(k,q)$ and $H_{\mu\nu}(q,k)$; 
their  most general Lorentz decomposition is given by 
\bea
-{\gb}_{\mu}(k,q) &=&  B_1(k,q) q_{\mu} + B_2(k,q) k_{\mu}.
\nonumber \\
-iH_{\mu\nu}(k,q) &=&  A_1(k,q) g_{\mu\nu}  + A_2(k,q) q_{\mu}q_{\nu} + A_3(k,q) k_{\mu}k_{\nu} 
+ A_4(k,q) q_{\mu}k_{\nu} +  A_5(k,q) k_{\mu}q_{\nu}
\nonumber \\
&&{}
\label{HGtens}
\eea
and from Eq.~(\ref{qH})
we obtain two constraints for the various form-factors, namely  
\bea
B_1(k,q) &=& A_1(k,q) + q^2 A_2(k,q) + (k \cdot q) A_4(k,q),
\nonumber\\
 B_2(k,q)&=& (k \cdot q) A_3(k,q) + q^2 A_5(k,q).
\label{con1}
\eea
Of course, since we work in the Landau gauge, due to the transversality of the gluon propagator,
the only relevant form factors are 
\bea
-\gb_\mu(k,q)&=&B_1(k,q)q_\mu,\nonumber \\
-iH_{\mu\nu}(k,q)&=& A_1(k,q)g_{\mu\nu}+A_2(k,q)q_\mu q_\nu.
\label{HGtensLG}
\eea

In the Landau gauge, the form factor
$B_1$ of Eq.~(\ref{HGtens}) is ultraviolet finite at one-loop, 
and therefore, no infinite renormalization constant needs to be introduced at that order;
of course, $B_2$ must be ultraviolet finite in all gauges, and to all orders, otherwise the
theory would be non-renormalizable. 
In order to obtain information about the ultraviolet behavior of $B_1$ beyond one-loop, 
one usually invokes the non-renormalization theorem of Taylor~\cite{Taylor:1971ff}, which states 
that for vanishing ghost momentum one has that 
\mbox{$B_1(q,-q)+ B_2(q,-q) =1$}, to all orders in perturbation theory.
Given that $B_2$ is finite to all orders (for any kinematic configuration), 
it follows that $B_1(-q,q)$ is also {\it finite} to all orders. 

Turns out that the vertex $\gb_\mu(k,q)$ has been studied on the lattice in the Landau gauge,  
for the Taylor kinematics, both for $SU(2)$~\cite{Cucchieri:2004sq} and $SU(3)$~\cite{Ilgenfritz:2006gp}.
According to these studies, $B_1(-q,q)$ deviates very mildly from 1 (the tree-level value). 
Even though the integration over the gluon momentum in the integrals of Eq.~(\ref{s4}) 
moves one away from the Taylor kinematics, the IR regime is well-represented,  
and we will approximate $B_1(k,q)$ by its tree-level value. 

On the other hand, to the best of our knowledge, 
the vertex  $H_{\mu\nu}(q,k)$ has not been studied on the lattice yet. 
Thus, the only constraints available are those coming from Eq.~(\ref{con1}); 
we will simply satisfy it by setting  $A_1(k,q)=  B_1(k,q)=1$ and \mbox{$A_2(k,q)=0$}. 

Then, under these approximations, the equations in (\ref{s4}) become
\bea
u(q^2) &=& \frac{g^2 C_{\rm {A}}}{d-1}\int_k \left[(d-2)+ \frac{(k \cdot q)^2}{k^2 q^2}\right]\Delta (k)  D(k+q),
\nonumber\\
L(q^2) &=& \frac{g^2 C_{\rm {A}}}{d-1}\int_k \left[1 - d \, \frac{(k \cdot q)^2}{k^2 q^2}\right]\Delta (k)  D(k+q).
\label{simple}
\eea

\subsection{Renormalization}

Now, as discussed in detail in~\cite{Aguilar:2009nf},  
the (unrenormalized) equations in (\ref{simple}) must be properly renormalized, {\it i.e.},  
in such a way as to preserve the validity  of the (BRST-induced) second identity in (\ref{ids}).
Specifically, 
the quantities $G(q^2)=u(q^2)$, $L(q^2)$, and $F(q^2)$
appearing in Eq.~(\ref{ids}) are unrenormalized 
(we have suppressed the corresponding subscript ``0'' for simplicity).
Note in fact that Eq.~(\ref{ids}) constrains the cutoff-dependence 
of the quantities involved; 
it is easy to recognize, for example,  by substituting into (\ref{simple}) and (\ref{SDgh})
tree-level expressions,
that  $F^{-1}(q^2)$ and  $u(q^2)$ have the same 
leading dependence on the UV cutoff $\Lambda$, namely [{\it viz}. (\ref{fx})]
\be
F^{-1}_{\chic{\mathrm{UV}}} (q^2) = u_{\chic{\mathrm{UV}}} (q^2) = 
\frac{9 \alpha_s}{16 \pi} \ln\left(\frac{\Lambda^2}{q^2}\right), 
\label{uvdiv}
\ee
while $L(q^2)$ is finite (independent of $\Lambda$).

Let us now denote by $Z_u$ the (yet unspecified) renormalization constant relating 
the bare and renormalized functions, $\Lambda_0^{\mu\nu}$ and $\Lambda^{\mu\nu}$, through
\be
g^{\mu\nu} + \Lambda_0^{\mu\nu}(q) = Z_u^{-1} [g^{\mu\nu} +\Lambda^{\mu\nu}(q)].
\label{Lamrel}
\ee
Note that the inclusion of the ``zeroth-order'' term $g^{\mu\nu}$ on both sides of (\ref{Lamrel}) is absolutely 
essential for the self-consistency of the entire renormalization procedure. To be sure, the $g^{\mu\nu}$ term appears naturally, 
given, for example, the form of the BQI in (\ref{twobqi});
indeed, the multiplicative renormalizability of (\ref{twobqi}) 
requires that the combination given in  (\ref{Lamrel}) should be renormalized as a whole.  

As already mentioned above, the origin of  Eq.~(\ref{ids}) 
is the BRST symmetry of the theory; in that sense, Eq.~(\ref{ids}) has the same origin 
as the Slavnov-Taylor identities. Therefore, just as happens with 
the Slavnov-Taylor identities, Eq.~(\ref{ids}) does not get deformed after renormalization.
Of course, the prototype example of such a situation are the Ward identities of QED; the requirement that 
the fundamental Ward identity $q^{\mu}\Gamma_{\mu} = S^{-1}(p+q)-S^{-1}(p)$ should retain the same 
form before and after renormalization leads to the well-known textbook relation $Z_1=Z_2$ 
between the corresponding renormalization constants.  
Similarly, for the case at hand, the renormalization must be carried out is such a way as to 
preserve the form Eq.~(\ref{ids}).  
Specifically, denoting by $Z_{c}$ the renormalization constant of the ghost dressing function, i.e.,
\be
Z_c(\Lambda^2, \mu^2) F_0^{-1} (q^2,\Lambda^2) = F^{-1}(q^2,\mu^2)\,,
\label{Zc}
\ee
and using the definition given in Eq.~(\ref{Lamrel}), it is clear that  
in order to preserve the relation~(\ref{ids}) after renormalization, we must impose that 
\be
Z_u = Z_{c}.
\label{renconst3}
\ee
As a result, one must renormalize Eq.~(\ref{SDgh}) using Eq.~(\ref{Zc}), and  Eq.~(\ref{simple})
using the relations 
\be 
Z_c(\Lambda^2, \mu^2)[1+u_0(q^2,\Lambda^2)+L_0(q^2,\Lambda^2)] = 1+u(q^2,\mu^2)+ L(q^2,\mu^2).
\label{Zcren}
\ee

To carry out the renormalization explicitly, 
let us introduce in addition  
 \bea
\Delta(q^2;\mu^2)&=& Z^{-1}_{A}(\mu^2)\Delta_0(q^2),\nonumber\\
g(\mu^2) &=& Z_{g}^{-1}(\mu^2) g_0,
\label{renconst}
\eea
and remember that, in the Landau gauge, due to Taylor's theorem,  the vertex renormalization is 1.   
Thus, 
after imposing the MOM renormalization condition $F(\mu^2)=1$, 
going to Euclidean space, setting $q^2=x$, $k^2=y$ and $\alpha_s=g^2/4\pi$, 
and implementing the standard angular approximation,  
one finds that the renormalized version of  Eq.~(\ref{SDgh}) reads 
\be
F^{-1}(x) = Z_c - \frac{\alpha_s C_{\rm {A}}}{16\pi} \left[
\frac{F(x)}{x}\int_{0}^{x}\!\!\! dy\  y \left(3 - \frac{y}{x}\right) \Delta(y) 
+ \int_{x}^{\infty}\!\!\! dy \left(3 - \frac{x}{y}\right)\Delta(y) F(y) 
\right],
\label{Fr}
\ee
where the renormalization constant  $Z_c$ is given by 
\be
Z_c = 1+ \frac{\alpha_s C_{\rm {A}}}{16\pi}  \left[
\frac{1}{\mu^2}\int_{0}^{\mu^2}\!\!\! dy  y \left(3 - \frac{y}{\mu^2}\right) \Delta(y) 
+ \int_{\mu^2}^{\infty}\!\!\! dy \left(3 - \frac{\mu^2}{y}\right)\Delta(y) F(y) 
\right].
\label{Zcexp}
\ee
Then, given Eq.~(\ref{Zcren}), we have that the renormalized version of Eq.~(\ref{simple}), under the same approximations, reads 
\bea
1+u(x) &=&  Z_c - \frac{\alpha_s C_{\rm {A}}}{16\pi}\left[
\frac{F(x)}{x}\int_{0}^{x}\!\!\! dy\  y \left(3 + \frac{y}{3x}\right) \Delta(y) 
+ \int_{x}^{\infty}\!\!\! dy \left(3 + \frac{x}{3y}\right)\Delta(y)F(y) 
\right],
\nonumber\\
L(x) &=&  \frac{\alpha_s C_{\rm {A}}}{12\pi} \left[
\frac{F(x)}{x^2}\int_{0}^{x}\!\!\! dy\ y^2 \Delta(y) 
+ x \int_{x}^{\infty}\!\!\! dy \frac{\Delta(y) F(y)}{y}
\right].
\label{LGr}
\eea
From this last equation it is easy to see ({\it e.g.}, 
by means of the change of variables $y = zx$) that if $\Delta$ and $F$ are IR finite, then $L(0) = 0$, 
as mentioned before~\cite{notf}. 
Note that  one cannot choose simultaneously the 
condition $u(\mu^2)=0$ once $F(\mu^2)=1$ has been imposed; indeed,  
given that $L(\mu^2)\neq 0$, such a choice would violate the identity of Eq.~(\ref{ids}).

\subsection{Numerical analysis}

\begin{figure}[!t]
\begin{center}
\includegraphics[scale=0.8]{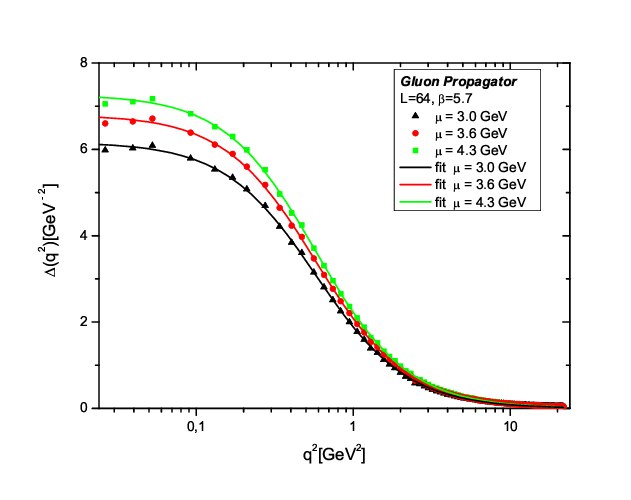}
\end{center}
\vspace{-1.0cm}
\caption{Lattice results for the gluon propagator renormalized at
 three different renormalization points: $\mu = 3.0 \,\mbox{GeV}$ (black curve), $\mu = 3.6 \,\mbox{GeV}$ (red curve) and $\mu = 4.3 \,\mbox{GeV}$ (green curve). 
We also show the corresponding fits using Eq.(\ref{fgluon}). 
The fitting parameters are: \mbox{$a=0.162\, \mbox{GeV}^2$}, \mbox{$b=0.367\, \mbox{GeV}^{-1}$} and \mbox{$c=1.5$} ($\mu = 3.0 \,\mbox{GeV}$); 
 \mbox{$a=0.147\, \mbox{GeV}^2$}, \mbox{$b=0.334\, \mbox{GeV}^{-1}$} and \mbox{$c=1.5$} (\mbox{$\mu = 3.6 \,\mbox{GeV}$}); \mbox{$a=0.137\, \mbox{GeV}^2$}, \mbox{$b=0.311\, \mbox{GeV}^{-1}$} 
and \mbox{$c=1.5$} (\mbox{$\mu = 4.3 \,\mbox{GeV}$}).} 
\label{fig1}
\end{figure}

\begin{figure}[!t]
\begin{minipage}[b]{0.5\linewidth}
\centering
\hspace{-1cm}
\includegraphics[scale=0.8]{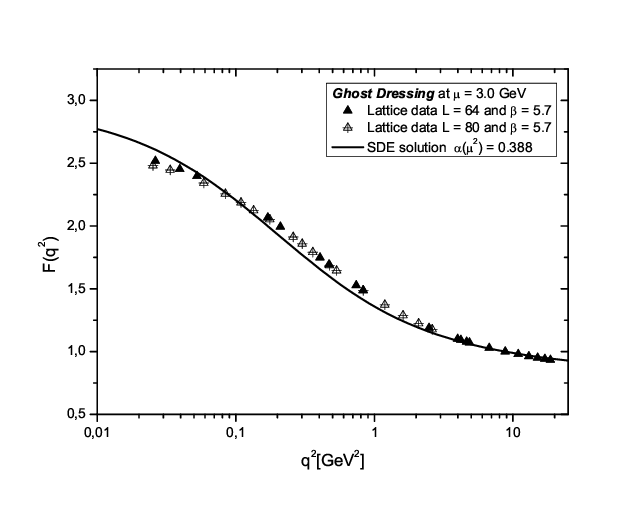}
\end{minipage}
\hspace{0.5cm}
\begin{minipage}[b]{0.45\linewidth}
\centering
\hspace{-2.0cm}
\includegraphics[scale=0.8]{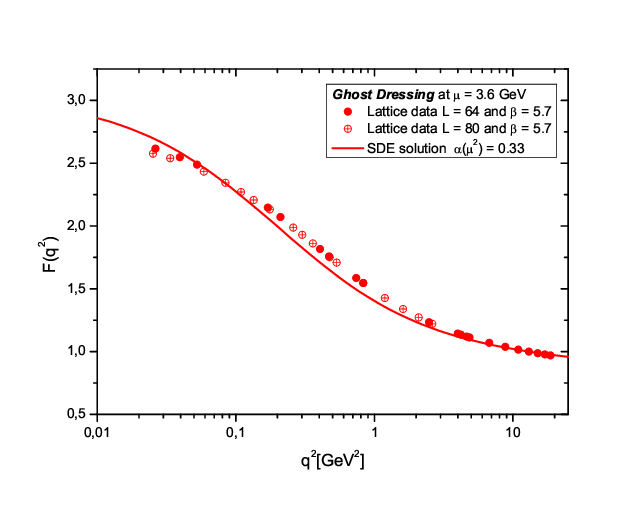}
\end{minipage}
\begin{minipage}[b]{0.5\linewidth}
\centering
\hspace{-1cm}
\includegraphics[scale=0.8]{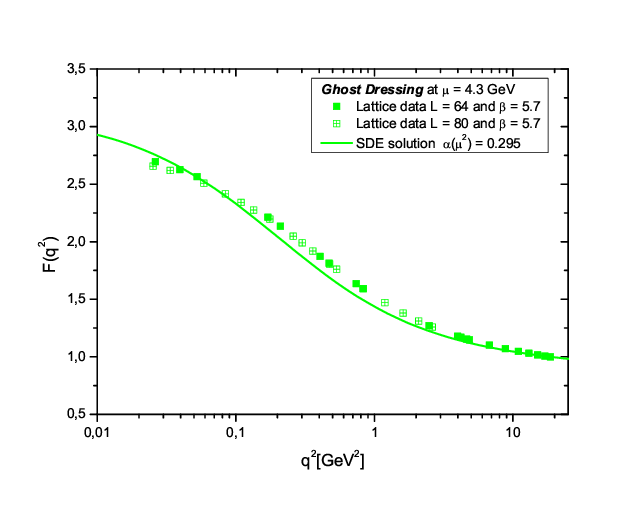}
\end{minipage}
\hspace{0.5cm}
\begin{minipage}[b]{0.45\linewidth}
\centering
\hspace{-2.0cm}
\includegraphics[scale=0.8]{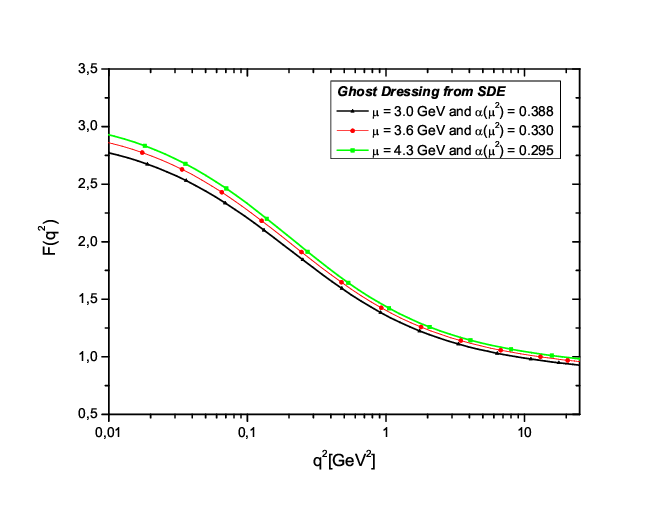}
\end{minipage}
\caption{{\it Top left panel}:
The ghost dressing function $F(q^2)$ obtained from SDE (black continuous line) compared to the lattice data of~\cite{Bogolubsky:2007ud} at $\mu = 3.0 \,\mbox{GeV}$. {\it Top right panel}: Same as in the previous panel but renormalized at \mbox{$\mu = 3.6 \,\mbox{GeV}$}.
{\it Bottom left panel}: Same as before but renormalized at \mbox{$\mu = 4.3 \,\mbox{GeV}$}. {\it Bottom right panel}: The SDE solutions for the three different renormalization points all together.}
\label{fig2}
\end{figure}

The starting point for our numerical analysis are the lattice results for 
the gluon propagator $\Delta(q^2)$ reported in~\cite{Bogolubsky:2007ud}. 
In order to eventually study the dependence of the KO function on the renormalization point, 
we would like to obtain the lattice data at different  renormalization points.
Since the gluon propagator is multiplicatively renormalizable, the relation~\cite{multren}
\be
\Delta(q^2,\mu^2)=\frac{\Delta(q^2,\nu^2)}{\mu^2\Delta(\mu^2,\nu^2)} \,, 
\label{ren_gl}
\ee
can be used to connect a set of points renormalized at $\mu$ with the corresponding set renormalized at $\nu$.
 Choosing the three 
different values $\mu=\{3.0,\ 3.6,\ 4.3\}\ \mathrm{GeV}$, we then obtain the three curves shown in Fig.~\ref{fig1}. 
In the range of available
momenta, a very accurate fit is provided by the expression
\be
\Delta(q^2)= \frac{1}{a+bq^{2c}},
\label{fgluon}
\ee
as shown by the continuous line in Fig.~\ref{fig1} (the values of the fitting parameters $a,\ b$, and $c$ are also reported there).

\begin{figure}[!t]
\begin{minipage}[b]{0.5\linewidth}
\centering
\hspace{-1cm}
\includegraphics[scale=0.8]{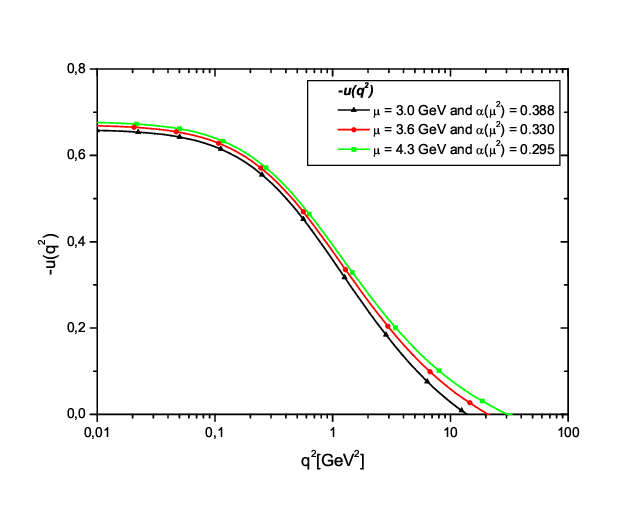}
\end{minipage}
\hspace{0.5cm}
\begin{minipage}[b]{0.45\linewidth}
\centering
\hspace{-2.0cm}
\includegraphics[scale=0.8]{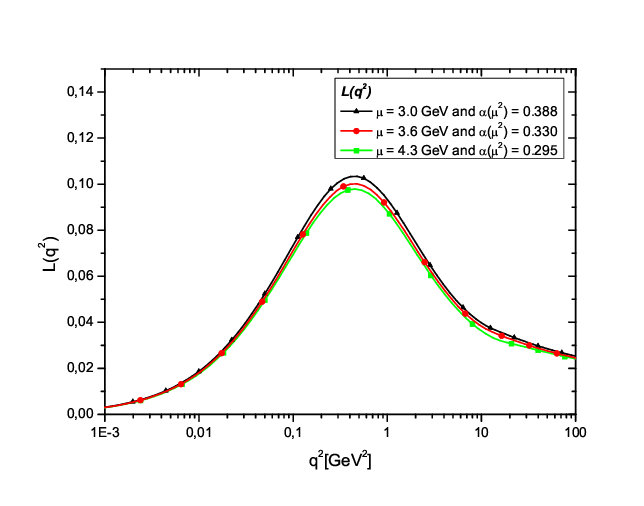}
\end{minipage}
\vspace{-1.5cm}
\caption{{\it Left panel}: $-u(q^2)$ determined from Eq.~(\ref{LGr}), using the solutions
for $\Delta(q^2)$ and $D(q^2)$ presented in Figs.~\ref{fig1} and~\ref{fig2} at the same renormalization points $\mu$. {\it Right panel}:  Same as in the previous panel but this time for $L(q^2)$. }
\label{fig3}
\end{figure}

The next step is to employ  the ghost SDE given in (\ref{Fr}) in order to 
deduce the appropriate values that one must use for the gauge coupling. 
To that end we will follow the following steps: (i) 
employing once again the relation (\ref{ren_gl}), with $\Delta \to D$, we generate from 
the lattice data on $F(q^2)$  
reported in~\cite{Bogolubsky:2007ud} the data sets for $F(q^2)$ corresponding to the 
renormalization points used previously for $\Delta$; 
(ii) using as input in (\ref{Fr}) 
the different sets of results obtained in the previous step for $\Delta$, we 
solve the integral equation (\ref{Fr}) numerically, thus determining $F(q^2;\mu^2)$;  
(iii) the values of $\alpha_s(\mu^2)$ are fixed by demanding 
that the solutions obtained in (ii) match the different lattice sets generated at step (i).

\begin{figure}[!t]
\begin{center}
\includegraphics[scale=0.8]{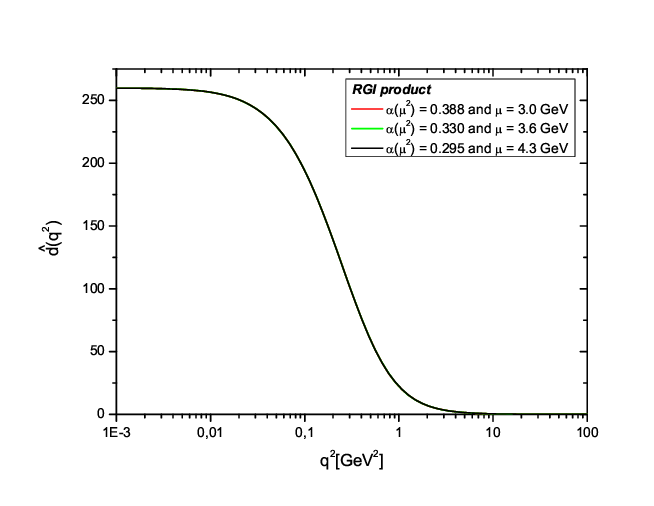}
\end{center}
\caption{The renormalization-group invariant product $\widehat{d}(q^2)$ obtained combining our results for $\Delta(q^2)$ and $G(q^2)$ according to Eq.~(\ref{rgi}).}
\label{fig4}
\end{figure}

The results of this procedure are displayed in Fig.~\ref{fig2}, where we show both the comparison between 
the lattice data and the solutions of Eq.~(\ref{Fr}) (first three panels), as well as the dependence on $\mu^2$ of these solutions (fourth panel). 
The couplings found are \mbox{$\alpha(\mu^2)=\{0.388,\ 0.330,\ 0.295\}$} for \mbox{$\mu=\{3.0,\ 3.6,\ 4.3\}\ \mathrm{GeV}$} respectively. 
Note that the values of $\alpha_s(\mu^2)$ obtained by this procedure are about 20$\%$ higher than those 
found from the two-loop MOM calculation of \cite{Boucaud:2005rm}. 

At this point, all necessary ingredient for determining the functions $u(q^2)$ and $L(q^2)$ are available. 
Substituting them into the corresponding equations given in~(\ref{LGr}), we obtain the solutions 
shown in Fig.~\ref{fig3}. Notice that $L(q^2)$ vanishes in the deep IR, as expected. 

A very stringent test of the quality of the obtained solutions can be devised by observing that, on formal grounds, the combination 
\be
\widehat{d}(q^2)= \frac{g^2(\mu^2) \Delta(q^2;\mu^2)}{\left[1+u(q^2;\mu^2)\right]^2},
\label{rgi}
\ee
constitutes a renormalization group invariant ({\it i.e.}, $\mu$-independent) quantity~\cite{Aguilar:2007ku}. 
Indeed, it is well-known that, 
due to the Abelian Ward identities satisfied by the PT-BFM Green's functions, the 
propagator $\widehat\Delta^{-1}(q^2)$ absorbs all Ê
the RG logs, exactly as happens in QED with the photon self-energy.
Specifically, if we define the renormalization constants 
of the gauge-coupling Êand the effective self-energy Êas
\bea
g(\mu^2) &=&Z_g^{-1}(\mu^2) g_0 ,\nonumber \\
\widehat\Delta(q^2;\mu^2) & = & \widehat{Z}^{-1/2}_A(\mu^2)\widehat{\Delta}_0(q^2), 
\label{conrendef}
\eea
then, since the renormalization constants above satisfy the QED-like relation 
\be
{Z}_{g} = {\widehat Z}^{-1/2}_{A}, Ê
\label{ptwi}
\ee
the product 
\be
{\widehat d}_0(q^2) = g^2_0 \widehat\Delta_0(q^2) = g^2 \widehat\Delta(q^2) = {\widehat d}(q^2), 
\label{ptrgi}
\ee
retains the same form before and after renormalization, {\it i.e.},, it 
forms a RG-invariant ($\mu$-independent) quantity~\cite{Cornwall:1982zr}.

In Fig.~\ref{fig4} we plot the combination above for the three different values of $\mu$ chosen; evidently the 
product of $g^2(\mu^2)$, $\Delta(q^2;\mu^2)$ and $[1+u(q^2;\mu^2)]^{-2}$ constructed from the solutions obtained is $\mu$-independent to an extremely high degree of accuracy.

\begin{figure}[!t]
\begin{minipage}[b]{0.5\linewidth}
\centering
\hspace{-1cm}
\includegraphics[scale=0.6]{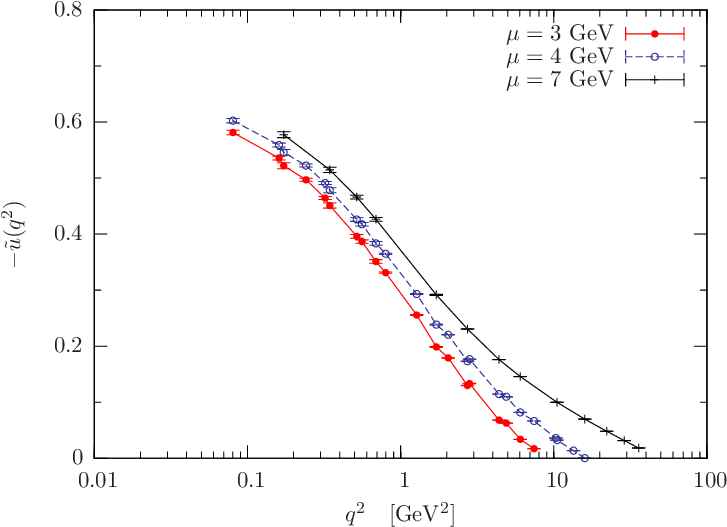}
\end{minipage}
\hspace{0.5cm}
\begin{minipage}[b]{0.45\linewidth}
\centering
\hspace{-2.0cm}
\includegraphics[scale=0.6]{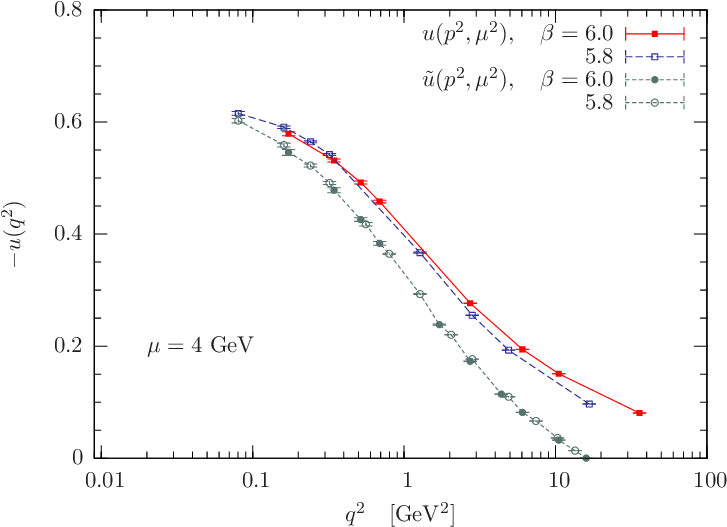}
\end{minipage}
\vspace{-1cm}
\caption{The determination of the KO function obtained in~\cite{Sternbeck:2006rd}.
{\it Left panel}: the function \mbox{$-\tilde u(p^2)=1- F^{-1}(p^2)$},  determined from Eq.~(\ref{ids}) by neglecting the $L(p^2)$ component, plotted at three different renormalization points. {\it Right panel}: Comparison between the function $-u(p^2)$ measured directly on the lattice and its asymptotic behavior $-\tilde u(p^2)$.}
\label{fig3-bis}
\begin{picture}(1,1)(0,0)
\rput(-5.5,8.8,0){\tiny{\gray Sternbeck hep-lat:0609016}}
\rput(2.55,8.8,0){\tiny{\gray Sternbeck hep-lat:0609016}}
\end{picture}
\end{figure}
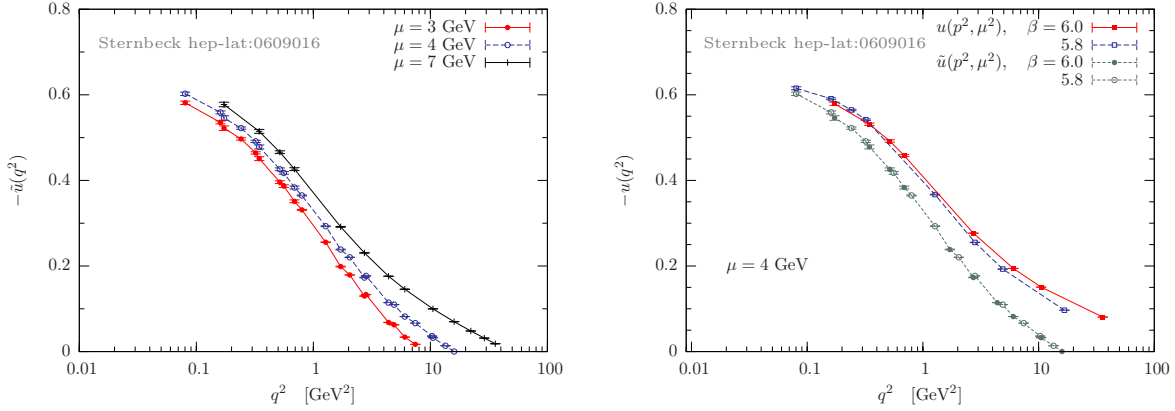

\begin{figure}[t]
\begin{center}
\includegraphics[scale=0.8]{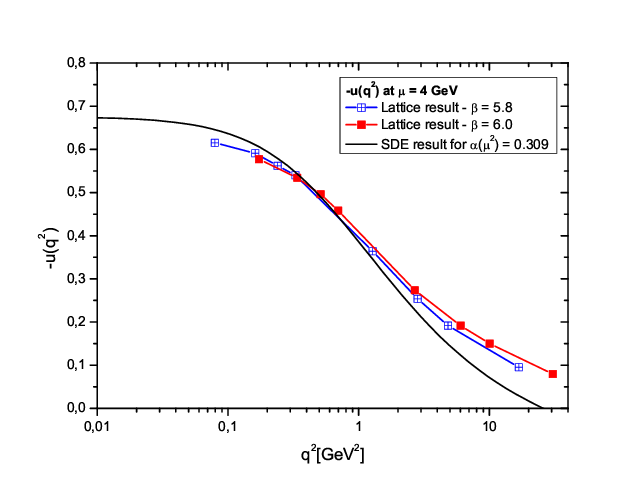}
\end{center}
\caption{\label{fig3-tris}
The KO function, $-u(q^2)$, obtained from the solution of Eq.~(\ref{LGr}) (black
continuous line) compared to the lattice data of ~\cite{Sternbeck:2006rd} at $\mu = 4\,\mbox{GeV}$.} 
\label{fig7}
\end{figure}

It is interesting to compare the curves plotted for $u(q^2)$ (Fig.~\ref{fig3} left panel) with those 
obtained by Sternbeck~\cite{Sternbeck:2006rd} (reproduced in Figs.~\ref{fig3-bis} and~\ref{fig3-tris}), where   
the function~(\ref{KO-1}) was studied in terms of Monte Carlo averages, and its asymptotic behavior was 
inferred from the identity~(\ref{ids}). In that case however the extrapolation in the deep IR region 
was problematic, due to a lack of knowledge of the function $L(q^2)$ [there denoted by $q^2v(q^2)$]; 
our analysis does not suffer from such a limitation, given that $L(q^2)$ is completely determined by its own equation.

One can see that the behavior is clearly the same encountered here (including the $\mu$-dependence); in addition we notice that the remarkable agreement found between the KO function extracted using our method and the direct calculation on the lattice, shows a posteriori that our tree-level approximations for the vertices appearing in the SDEs~(\ref{s4}) is indeed justified.

\begin{figure}[!t]
\begin{minipage}[b]{0.5\linewidth}
\centering
\hspace{-1cm}
\includegraphics[scale=0.8]{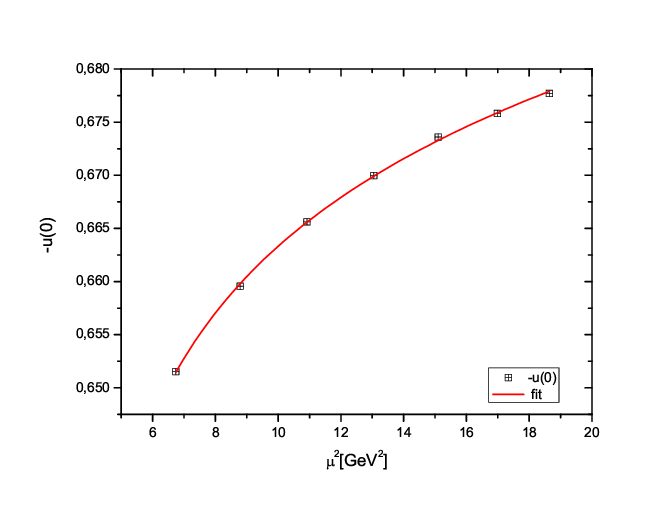}
\end{minipage}
\hspace{0.5cm}
\begin{minipage}[b]{0.45\linewidth}
\centering
\hspace{-2.0cm}
\includegraphics[scale=0.8]{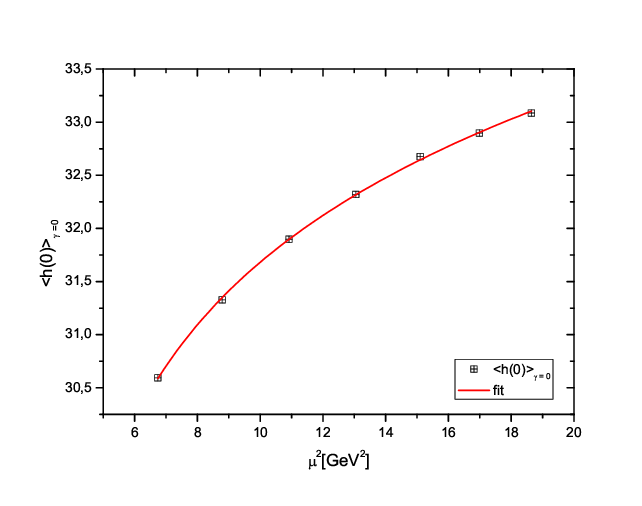}
\end{minipage}
\vspace{-1.5cm}
\caption{{\it Left panel}: The dependence of the KO parameter $u(0)$ on the renormalization
point $\mu$.  {\it Right panel}:  Same as in the previous panel but for the horizon function. 
In both cases  the continuous red line represents the fit given by Eq.~(\ref{fitg0}) with   $a_1=0.633$, $b_1=3.57$, $c_1=0.025$ and  $a_2=28.61$, $b_2=3.25$, $c_2=0.05$ for -$u(0)$ and $\langle h(0)\rangle_{\gamma=0}$ respectively.}
\label{fig5}
\end{figure}

Now,  the important  point to emphasize  is  that the
function  $u(q^2)$ is {\it  not} a  $\mu$-independent quantity; in fact, as we have established (within the MOM scheme)
its $\mu$-dependence is exactly what is needed in order to enforce the $\mu$-independence of 
the RG-invariant expression given in (\ref{rgi}).
In fact, its value at $q^2=0$, {\it i.e.}, the KO parameter $u(0)$, {\it depends} on
the renormalization point. 
Notice that in the case  of an IR divergent
ghost dressing function the possible  $\mu$-dependence would be inconsequential, 
since, due to the   identity~(\ref{ids}), $u(0)=-1$ irrespectively of  the value  of $\mu$ chosen. 
Evidently, the situation  is different 
in the  case of  an IR  finite ghost  dressing function,
since  $u(0)$ acquires  a non-trivial  dependence on  the renormalization
scale. This dependence is plotted on the left panel of Fig.~\ref{fig5} for values 
of $\mu$ varying between 2.6 and 4.3 GeV (due to the limited number of UV lattice points of our data), 
where we see that $-u(0)$ varies in the interval [0.65, 0.68], in good agreement with the prediction~(\ref{new-pred}).
The dependence of $u(0)$ on $\mu$ appears to be moderate, probably due to the rather narrow region 
of allowed $\mu$ values considered. A more detailed study with data sets extending deeper in the UV 
should allow one to explore the full extent of this dependence.

On the right panel Fig.~\ref{fig5} we plot finally the $\mu$-dependence of the horizon function (\ref{res-1}).
Notice that both dependencies can be fitted with a function that is characteristic of a phase transition, namely
\bea
-u(0)&=&a_1(\mu^2-b_1)^{c_1} , \\ \nonumber
\langle h(0)\rangle_{\gamma=0}&=&a_2(\mu^2-b_2)^{c_2} ,
\label{fitg0}
\eea
as shown by the continuous red curves appearing in  Fig.~\ref{fig5} (the values of the fitting parameter are also reported there).

\section{\label{concl}Conclusions}

In this work we have presented an indirect determination of the 
KO function from recent lattice data on the behavior of
the   QCD  gluon   and  ghost   propagators~\cite{Bogolubsky:2007ud,Cucchieri:2007md} 
in the Landau gauge. The results obtained are in very good agreement with the 
original study of the same quantity presented in ~\cite{Sternbeck:2006rd}. 

Of particular interest is the observed dependence of the KO function, 
and particular of its infrared value $u(0)$,  
on  the renormalization point $\mu$ chosen  within the MOM scheme.  
The $\mu$-dependence of $u(0)$ within the latter scheme, 
mild as it may seem at first sight, is definitely there, as one would expect, given that the 
KO function $u(q^2)$ is {\it not} a renormalization-group  invariant quantity, {\it i.e.}, it is not
intrinsically  $\mu$-independent. 
In fact, the observed $\mu$- dependence 
is really sizeable when contrasted with the impressive absence of any $\mu$-dependence displayed 
by a genuinely $\mu$-independent quantity given in Eq.(\ref{rgi})
which was computed using exactly the same sets of lattice data.   
We hope that the present work will contribute to the study  
of the possible effects that renormalization may have on the 
quantitative predictions of the KO formalism. 

\begin{figure}[!t]
\begin{center}
\includegraphics[scale=0.8]{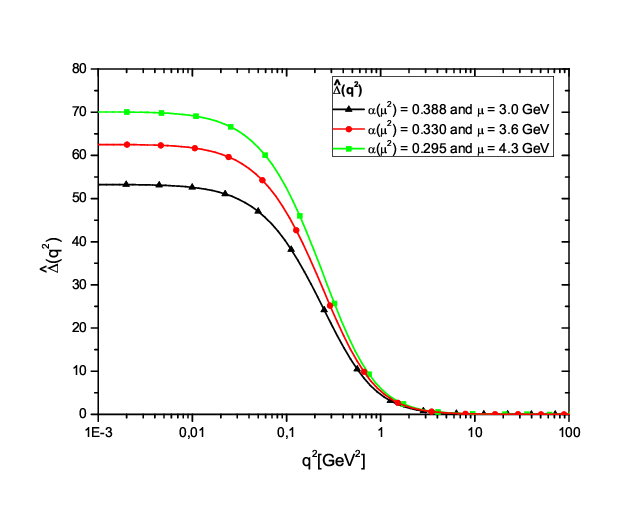}
\end{center}
\vspace{-1.3cm}
\caption{ The gluon propagator $\widehat{\Delta}(q^2)$ of the BFM, renormalized at
three different points: \mbox{$\mu = 3.0 \,\mbox{GeV}$} (black curve), \mbox{$\mu = 3.6 \,\mbox{GeV}$} (red curve) and \mbox{$\mu = 4.3 \,\mbox{GeV}$ (green curve)}.}
\label{fig6}
\end{figure}

Let us take a closer look at the background-quantum identity given in Eq.~(\ref{DeltaBQI}), which 
relates the conventional gluon propagator $\Delta$ with the gluon propagator $\widehat{\Delta}$
of the BFM.  Eq.~(\ref{DeltaBQI}) assumes that the  
corresponding gauge-fixing parameters, namely the $\xi$ of the $R_{\xi}$ 
and the $\xi_{Q}$ used in the BFM to gauge-fix the quantum fields appearing inside the loops, 
are equal ($\xi=\xi_{Q}$). In the Landau gauge, $\xi=\xi_{Q}=0$, due to the central equality of 
Eq.~(\ref{KO-G}), we have that 
\be
u(q^2) =  \sqrt{\frac{\Delta(q^2)}{\widehat{\Delta}(q^2)}} -1 .
\label{ulatt2}
\ee
Interestingly enough, this simple formula expresses the KO function in terms 
of two gluon propagators calculated in the Landau gauge of two very distinct 
gauge-fixing schemes, with no direct reference to the ghost sector of the theory.
This observation opens up the possibility 
of deducing the structure of the KO function using an entirely different, and completely novel, approach.
Specifically, one may envisage a lattice simulation of 
$\widehat{\Delta}$~\cite{Dashen:1980vm}; then, $u(q^2)$ may be obtained from (\ref{ulatt2}) by simply  
forming the ratio of the two gluon propagators. 
Given that $\Delta(0)$ is found to be finite on the lattice~\cite{Bogolubsky:2007ud,Cucchieri:2007md}, it is clear that,
in order for the standard KO criterion to 
be satisfied ({\it i.e.}, $u(0)=-1$), $\widehat{\Delta}$ must diverge in the IR.  
Needless to say, we consider such a scenario highly unlikely. 
What is far more likely to happen, in our opinion, is to find a perfectly finite and well-behaved 
$\widehat{\Delta}$, which in the deep IR  
will be about an order of magnitude larger than $\Delta(0)$, furnishing a value $u(0)\sim -0.6$, namely   
what we have found in our analysis. In fact, one may turn the argument around: 
combining the results of this article with the lattice data for $\Delta$~\cite{Cucchieri:2007md,Bogolubsky:2007ud}, 
one may use (\ref{ulatt2}) to predict the outcome of the lattice simulation for $\widehat{\Delta}$; 
our prediction for the case of $SU(3)$ is shown in Fig.~\ref{fig6}.

\acknowledgments

We thank Prof. K-I. Kondo for bringing to our attention reference~\cite{Sternbeck:2006rd}.
The research of J.~P. is supported by the European FEDER and  Spanish MICINN under grant FPA2008-02878, and the Fundaci\'on General of the UV.


\begin{thebibliography}{99}
\bibitem{KOdef} In this paper we will refer to $u(q^2)$ as the ``KO function''; 
$u(0)$ is known in the literature as the  ``KO parameter''.
\bibitem{KOpred} The  KO scenario predicts a gluon  propagator that is
less  divergent  than  the  tree-level  expression;  this,  evidently,
encompasses  the IR-finite  gluon propagator  as a  special  case, even
though, up  until recently,  the focus had  been placed rather  on the
``vanishing'' solutions,  given that they  satisfy simultaneously both
the KO and GZ requirements.
\bibitem{notf}
 Note that the finiteness of $\Delta$ and $F$ is not a necessary condition for $L(0)$ to vanish; 
for example, one may obtain $L(0) = 0$ using $\Delta(y) \sim y^{a}$ and  $F(x)\sim x^{b}$, provided that $a+b>-1$.
\end{thebibliography}
\end{document}